\documentclass[
reprint,
superscriptaddress,
 amsmath,amssymb,
 aip, rsi, numerical
]{revtex4-1}

\usepackage{graphicx}
\usepackage{dcolumn}
\usepackage{bm}
\usepackage{natbib}
\usepackage{siunitx}
\usepackage{textcomp}

\begin{document}

\preprint{APS/123-QED}

\title{Surface trap with dc-tunable ion-electrode distance}

\author{Da An}
\author{Clemens Matthiesen}
\author{Ahmed Abdelrahman}
\author{Maya Berlin-Udi}
\author{Dylan Gorman}
\author{S\"{o}nke M\"{o}ller}
\author{Erik Urban}
\author{Hartmut H\"{a}ffner}

\affiliation{%
Department of Physics, University of California, Berkeley, California 94720, USA
}%

\date{\today}

\begin{abstract}
We describe the design, fabrication, and operation of a novel surface-electrode Paul trap that produces a radio-frequency-null along the axis perpendicular to the trap surface. This arrangement enables control of the vertical trapping potential and consequentially the ion-electrode distance via dc-electrodes only. We demonstrate confinement of single $^{40}$Ca$^+$ ions at heights between 50~{\textmu}m and 300~{\textmu}m above planar copper-coated aluminium electrodes. We investigate micromotion in the vertical direction and show cooling of both the planar and vertical motional modes into the ground state. This trap architecture provides a platform for precision electric-field noise detection, trapping of vertical ion strings without excess micromotion, and may have applications for scalable quantum computers with surface ion traps. 
\end{abstract}

\maketitle


\section{\label{sec:intro}Introduction}
Long coherence times and precise control of electronic states place trapped ions at the forefront of research on quantum sensing, simulation, and computation \cite{Haeffner2008review}. Radio-frequency (rf) surface ion traps, taking advantage of microfabrication technology, allow for miniaturization of trap designs \cite{Seidelin2006} and the realization of more complex trap architectures \cite{Chiaverini2005, Kielpinski2002}. For instance, shuttling ions between trapping regions and through junctions \cite{Shu2014HeatingTrap, Wright2013, Moehring2011} and on-chip integration of ion addressing beams have been demonstrated \cite{Mehta2016}. Recent efforts have been directed at mitigating the disadvantages of surface traps, such as low trap depths, limited optical access \cite{Maunz2016}, and susceptibility to anomalous heating due to proximity of the ion to the trap surface \cite{Brownnutt2015}. Studies with surface ion traps have, to this point, mostly relied on the five-wire trap design \cite{Chiaverini2005, Pearson2006}, where the trap axis without rf confinement is oriented parallel to the surface. Strings of ions are then typically trapped along this axis and confined with dc fields. Changing the ion height, defined as the perpendicular ion-surface distance, \textit{in-situ} while also avoiding excess micromotion has been realized by application of additional rf signals on the trap chip. \cite{Kim2010b, VanDevender2010, Boldin2018MeasuringSeparation}.

Here, we introduce a novel surface trap design, the four-rf-electrode trap, where rf fields provide confinement in the trap plane, but cancel along the vertical axis which is perpendicular to the trap surface. The ion trapping height is controlled with dc fields, and we demonstrate trapping at heights ranging from 50 to 300~{\textmu}m. Full control of the vertical potential with dc voltages permits trapping of vertical ion strings and simplifies vertical shuttling, with immediate applications in sensing and quantum control.

\section{\label{sec:trapdesignfab}Trap Design and Fabrication}
\begin{figure*}[t!]
    \centering
    \includegraphics[scale=1]{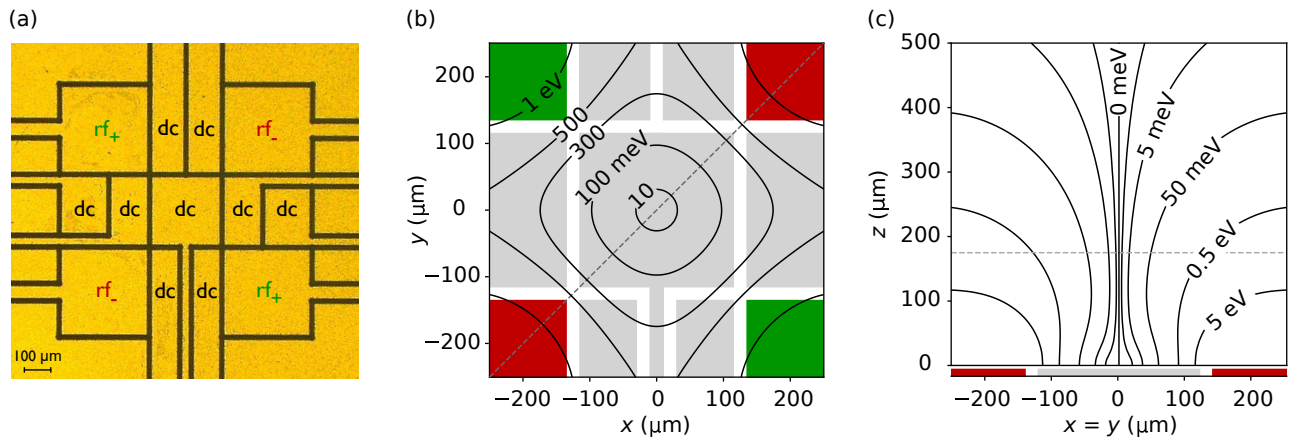}
    \caption{(a) False-color microscope image of the microfabricated four-rf-electrode trap. Static voltages are applied to electrodes labeled dc, and rf electrodes are labeled $\mathrm{rf_{\pm}}$ according to the signal phase. (b) Top-down view of $xy$-plane rf pseudopotentials at a height of $z=175$~{\textmu}m, with electrode geometry shown for reference. (c) Side view of rf potentials. The grey dashed lines in (b), (c) denote the position of the cross-sectional slice in (c), (b), respectively.}
    \label{fig:trap}
\end{figure*}

Our trap design takes inspiration from the cross-sectional cut of a conventional four-rod Paul trap \cite{Paul1990a}. Figure \ref{fig:trap}(a) shows a false-color photograph of the trap we fabricated and used in this report. Four rf electrodes measuring 290 $\times$ 290~{\textmu}$\mathrm{m}^{2}$ each are centered on the corners of a 560 $\times$ 560~{\textmu}$\mathrm{m}^{2}$ square while nine dc electrodes are sized and positioned to optimize control of static dipole and quadrupole fields. All other areas on the chip are grounded. 20-{\textmu}m-wide leads are attached to the rectangular dc electrodes to route voltages from off-chip sources, while leads to the rf electrodes are 40~{\textmu}m wide.  All electrodes are separated by trenches of 20-{\textmu}m width and 50-{\textmu}m depth. The electrodes are  patterned by laser-etching ultraviolet fused silica (performed by Translume, Ann Arbor, MI, United States). Subsequent layers of titanium (Ti), aluminum (Al), and copper (Cu) are deposited onto the substrate at an angle of \ang{60} to the surface normal using electron-beam physical vapor deposition, with respective thicknesses of 15~nm, 500~nm, and 30~nm. This process is then repeated at an angle of -\ang{60}, coating both the top surface of the electrodes and the upper walls of the trenches, thus shielding the ion from stray fields on the surface of the dielectric substrate material. Ti is used as a sticking layer, Al acts as the primary conducting material, and Cu protects the Al surface from rapid oxidation.

The four-rf-electrode trap can be operated in two configurations: as a point trap or as a vertical-linear trap. In the point trap configuration, all four rf electrodes are driven with the same amplitude and phase, generating three-dimensional rf confinement with a single rf-null trapping point. In the vertical-linear configuration, we employ an out-of-phase rf drive where one diagonal pair of rf electrodes ($\mathrm{rf_{+}}$) is driven with a sinusoidal voltage, while the other diagonal pair ($\mathrm{rf_{-}}$) is driven with a voltage of the same amplitude, but opposite phase. We have trapped single ions in both configurations, here we will focus on the operation of the four-rf-electrode trap in the vertical-linear configuration.

The out-of-phase rf drive of the vertical-linear trap produces an effective potential, or pseudopotential \cite{Leibfried2003}, originating from time-averaged rf fields, as shown in Fig. \ref{fig:trap} (b) and (c). The calculation of the potential assumes a strictly two-dimensional trap geometry, which can be solved analytically \cite{Wesenberg2008} taking into account each electrode's solid-angle with respect to the ion location. Our simulations assume an rf drive frequency $\Omega_{\mathrm{rf}} = 2\pi\times$18~MHz and a peak-peak voltage $V_{\mathrm{pp}}= 200$~V. Panel (b) displays the rf pseudopotential lines in the $xy$-plane at a height of 175~{\textmu}m, showing quadrupole confinement with an rf null at the trap center, $x=y=0$. The trap design is shown in the background for reference. Fig.~\ref{fig:trap} (c) shows a cross-section of the pseudopotential in the plane defined by the dashed grey line in panel (b) and the vertical $z$-axis. Due to the symmetry of the rf-electrode pairs, the rf fields fully cancel along the vertical $z$-axis at $x = y = 0$, creating an rf-null axis perpendicular to the surface. Confinement in this direction can be achieved with dc potentials, which should allow tuning the trapping location along the vertical axis without introducing excess micromotion.

\section{\label{sec:trapparams}Trap Operation}
We operate the trap in an ultra-high vacuum chamber at a pressure $<1\times 10^{-10}$~mbar. An rf voltage at a frequency of $\Omega_{\mathrm{rf}} = 2\pi\times$18.1~MHz is generated with a signal generator (Rohde \& Schwartz SMB 100A), amplified (mini-circuits ZHL-5W-1) and applied to the rf electrodes through an inductively coupled toroidal half-wave resonator. A beam of calcium atoms is propagated through the trapping region from a resistively heated oven filled with macro-granules of calcium (Ca). Neutral $^{40}$Ca atoms are selectively ionized in a two-step process using laser light at 422 and 375 nm. 
A pair of red-detuned 397-nm and 866-nm beams addresses the $\lambda$-system consisting of the electronic states $4^{2}S_{1/2}-4^{2}P_{1/2}-3^{2}D_{3/2}$ and cools the ion motion. Ion fluorescence at 397~nm is detected by a photomultiplier tube and used for determining whether the ions is in the $4^S_{1/2}$ or the $3^D_{5/2}$ state. In order to cool both planar and vertical motional modes the 397-nm beam propagates in the plane parallel to the trap surface, while the 866-nm beam is near-vertical. A 729-nm beam, which addresses the dipole-forbidden transition $4^{2}S_{1/2}-3^{2}D_{5/2}$ connecting our qubit states, is used for ion spectroscopy and sideband cooling. It can be switched between two propagation directions, one in the trap plane and one vertical to the trap surface.

With this configuration, we have demonstrated trapping of single ions at heights between 50~{\textmu}m and 300~{\textmu}m. The experimentally limiting factor for operation at both extremes is the high voltage required on the rf electrodes. At low trapping heights the trap harmonicity also decreases, resulting in ion lifetimes decreasing from several hours at 110~{\textmu}m height to several minutes at 50~{\textmu}m for the same secular frequencies. Typical secular trap frequencies range from $\omega_z = 2\pi\times0.4$ MHz to $2\pi\times1.2$~MHz and $\omega_x, \omega_y = 2\pi\times0.6$~MHz to $2\pi\times2.0$~MHz. The degeneracy between the two planar modes is lifted by applying a suitable dc quadrupole, resulting in a typical splitting of $|\omega_{x}-\omega_{y}|/ \omega_{x} \sim 0.1$. By changing the dc quadrupole values we also control the orientation of the principal axes \cite{Schindler2015}. This ability is important as it allows us to slightly tilt the vertical mode into the trap plane, thus creating non-zero projection with the in-plane 397-nm laser which improves cooling of the near-vertical mode.

\begin{figure}[t!]
    \centering
    \includegraphics[scale=1]{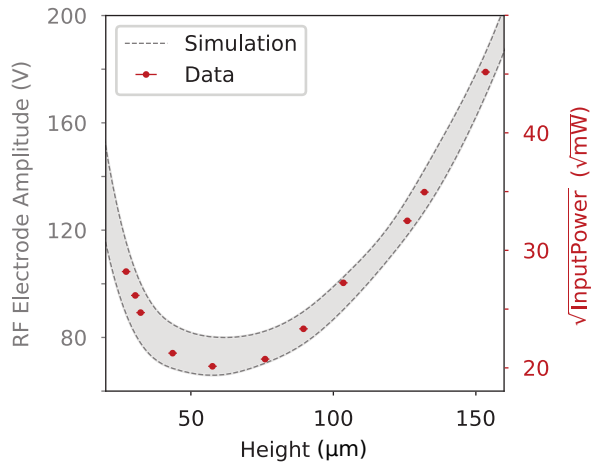}
    \caption{Rf voltages required to achieve a planar secular frequency of $2\pi\times1$~MHz as a function of the ion height. The grey band indicates the simulated voltages for vertical tilt angles of \ang{0} (lower limit) and \ang{4} (upper limit). Red dots show the square root of the power output from the rf signal generator applied to the rf resonator-trap system. Error bars correspond to one s.d. uncertainty in the ion height.}
    \label{fig:rfpower}
\end{figure}

While there is no rf confinement along the vertical axis, the rf field strength in the plane depends on the distance to the surface. This effect can be seen in Fig.~\ref{fig:trap}(c) where the equipotential lines are closest together around 110-{\textmu}m height. Experimentally, we observe this dependence by trapping a single ion at various ion-electrode distances and comparing the rf-power needed to create a planar confinement corresponding to a fixed secular frequency, $2\pi\times1$~MHz in this case.
Data and simulation of our applied rf power are displayed in Fig. \ref{fig:rfpower}. The blue shaded region indicates the simulated rf voltage amplitude, with the lower and upper limits determined by a vertical tilt angle of \ang{0} and \ang{4}, respectively, which are typical parameters in our measurements. The data points (red dots) show the square root of the power output from the rf signal generator, which is routed, via the amplifier, to the rf resonator-trap system. The ion trapping height is determined experimentally by maximizing ion fluorescence in the vertical direction and then translating the $\sim 5$-{\textmu}m wide 397-nm beam until it scatters from the center of the trap, taking care to avoid backlash of the motorized translation stage. The uncertainty in the ion-surface distance is estimated at $\pm 1.5$~{\textmu}m, given by the precision of repeated measurements. Overall, data and simulation agree well.

\begin{figure*}[t!]
    \centering
    \includegraphics[scale=1]{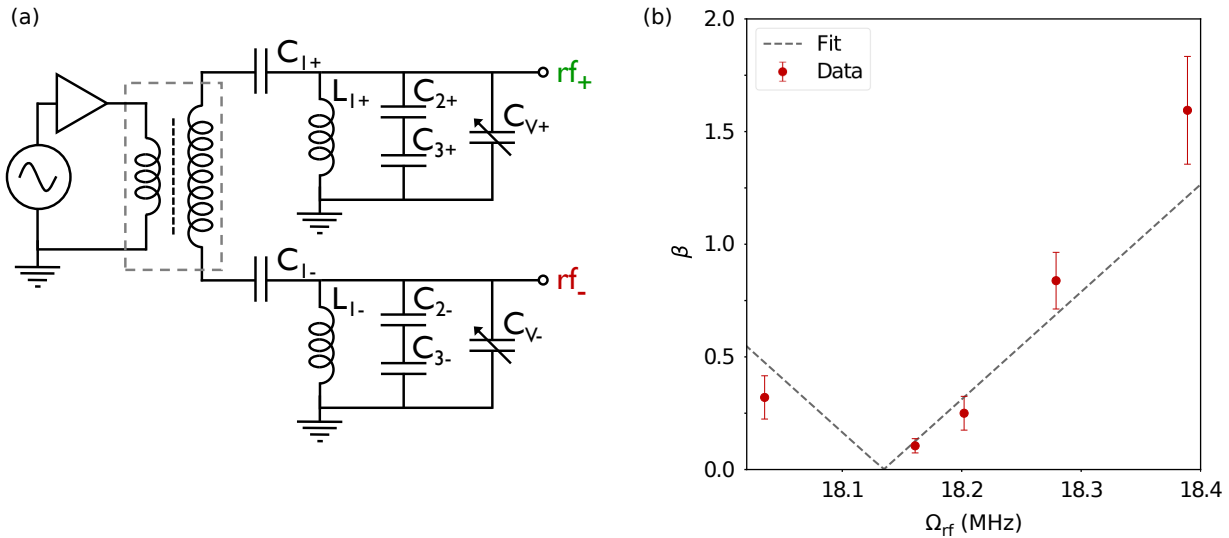}
    \caption{(a) Circuit schematic for the implementation of the out-of-phase rf drive. The boxed region indicates a toroidal transformer with a powdered iron core. Each rf signal arm is equipped with a high-pass filter ($C_{1\pm}$, $L_{1\pm}$), a capacitive divider ($C_{2\pm}$, $C_{3\pm}$), and tuning capacitor ($C_{V\pm}$). (b) Measured micromotion modulation index $\beta$ at a trapping height of 200~{\textmu}m. The rf drive frequency, $\Omega_{\mathrm{rf}}$ reflects the resonance of the rf circuit which is tuned via the capacitors $C_{V\pm}$. Here we change only one of the two tunable capacitors and the dashed line shows the expected linear dependence of the modulation index on the circuit resonance.}
    \label{fig:micromotion}
\end{figure*}

\subsection{\label{sec:micromotion}Resonator Circuit and Micromotion Compensation}
Generating perfectly canceling rf fields along the vertical axis of our trap relies on precise matching of the signal on the two diagonal pairs of rf electrodes. Ideally, they are at the same frequency and amplitude, but of opposite phase. However, any mismatch in those parameters, for instance due to fabrication imperfections or slight differences in the capacitance and inductance of wiring, leads to a residual rf field at the trapping location, and consequently, micromotion in the vertical direction. In a real device, such a mismatch can easily occur and is observed in our setup. In the following we describe the circuit used to deliver the rf voltages to the trap electrodes in more detail and show how to reduce vertical micromotion.

Fig. \ref{fig:micromotion} shows the circuitry for passively routing out-of-phase rf signals onto separate electrodes. Rf signals from the source are amplified and linked to the trap electrodes via a toroidal transformer, indicated by the boxed region in panel (a). The right-hand side of the transformer forms a resonant circuit with the cumulative capacitance to ground, which is dominated by the rf$_{\pm}$ electrode wiring. This resonator circuit matches the real 50-$\Omega$ resistance from the signal generator to the imaginary impedance of the trap electrodes. The two sides of the transformer output split the following circuitry into two arms with nominally equal amplitude, frequency, and opposite phase. Each arm contains a high-pass filter ($C_{1\pm}=2$~nF and $L_{1\pm}=100$~mH), where the inductors, $L_{1\pm}$, act as ground references for the rf signal. The amplitude and phase mismatch between the two rf arms can be measured with the capacitive dividers consisting of $C_{2\pm} = 2$~pF and $C_{3\pm} = 100$~pF. To correct for discrepancies, tunable capacitors, $C_{V\pm}$, with a range of 2 - 7~pF are added in parallel to the trap electrodes. These allow us to tweak the amplitude, frequency, and phase of the resonant circuit, thus reducing the mismatch of the two rf arms.

Imperfect cancellation of rf fields along the vertical $z$-axis manifests on the ion as excess micromotion in the $z$-direction. Micromotion, in turn, provides a much more relevant and sensitive measurement of differences in the resonator arms as compared to the capacitive divider signal. By tuning the capacitor on one arm, we can measure changes in the ratio of the first order micromotion sideband Rabi frequency, $\Omega_1$, and the Rabi frequency of the direct carrier transition, $\Omega_0$. Using electrostatic and lumped circuit simulations we find the modulation index, $\beta \approx 2 \Omega_1 / \Omega_0$ should depend to first order linearly on the the resonance frequency and theoretically vanishes when both arms match amplitudes and phases perfectly \cite{Keller2015PreciseClocks}. Experimental constraints, such as a finite resonator quality factor,  naturally limit how closely we can approximate the perfect case. In our case, cf. Fig.~\ref{fig:micromotion}, the modulation index could be reduced to $\beta \approx 0.1$, a considerable improvement over the un-optimized modulation index, $\beta \approx 1.5$. We have repeated the procedure for vertical micromotion compensation with two different resonant circuits and found similar behaviour for both.

\subsection{\label{sec:cohops}Ground state cooling}
\begin{figure}[b!]
    \centering
    \includegraphics[scale=1]{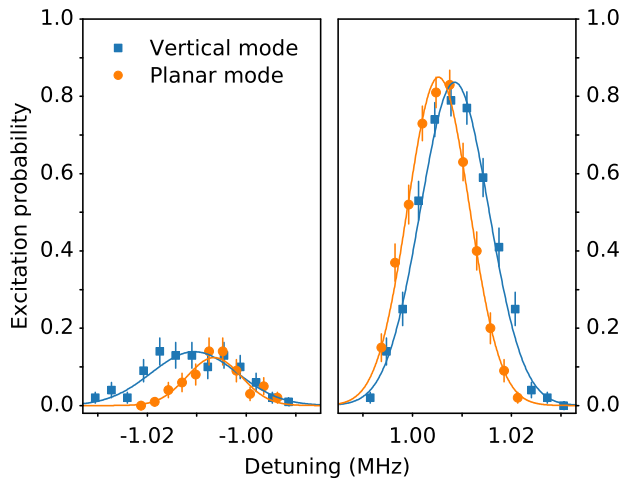}
    \caption{Demonstration of ground state cooling of vertical and planar modes at a trapping height of 114~{\textmu}m. The left (right) panel displays the red (blue) sideband excitation probability. Data for the vertical (planar) mode are shown as blue squares (orange circles), solid lines are Gaussian fits to the data. The mean phonon numbers for the planar and vertical modes are 0.17(3) and 0.20(3) quanta at $2\pi \times 1$~MHz, respectively. Error bars are based on binomial statistics for 100 measurement repetitions and represent one s.d. uncertainty.}
    \label{fig:sidebands}
\end{figure}
Finally, we demonstrate ground state cooling of the ion's motion in this four-rf-electrode trap which is relevant for high-fidelity sensing and quantum control applications. Both vertical and planar motional modes are of interest for future experiments, so we perform two sets of measurements, one for the planar modes and, switching the direction of the 729-nm laser to be perpendicular to the trap surface, one for the vertical mode. We operate at a trapping height of 114~{\textmu}m and tune the secular frequency to $\sim 2\pi \times 1$~MHz in both cases. During the cooling sequence we Doppler-cool on the $4^{2}S_{1/2}-4^{2}P_{1/2}$ transition for 2~ms, then perform sideband-cooling using the first-order secular sideband of our qubit transition for six 1-ms cycles. A measurement of the ratio of Rabi frequencies for the red and blue sidebands can be used to calculate the mean phonon number for the motional mode~\cite{Leibfried2003}. Figure~\ref{fig:sidebands} displays measurements of the red (left panel) and blue (right panel) secular sidebands for the planar (orange circles) and vertical (blue squares) modes, together with Gaussian fits as solid lines. We obtain mean mode occupations of $\bar{n}_\mathrm{planar} = 0.17(3)$ and $\bar{n}_\mathrm{vertical} = 0.20(3)$ phonons, corresponding to a ground state occupation of 0.85 and 0.83, respectively. The achievable ground state occupation at this trapping height is limited by the intensity of the 729-nm light used for sideband cooling in conjunction with heating rates of several hundred quanta per second. 

\section{Conclusions}
The ability to tune the ion trapping height by changing the dc fields opens up new opportunities for sensing and quantum information applications. Of immediate interest are studies of electric field surface noise as a function of ion-surface distance. The distance scaling of the spectral noise density, $S_{E}$, is an important indicator for the type and origin of surface noise \cite{Brownnutt2015}. Very recently, two studies independently showed a d$^{-4}$ power-law scaling for an electroplated gold surface trap \cite{Boldin2018MeasuringSeparation} and a sputtered niobium trap \cite{Sedlacek2018DistanceTrap}. We aim to measure the scaling in our Al-Cu four-rf-electrode trap, which allows tuning the ion-electrode distance continuously from 50~{\textmu}m to 300~{\textmu}m. Access to motional modes both vertical and horizontal to the electrode surface should deliver further information as to the characteristics of the surface noise.

Beyond studies of the environment noise, we foresee applications in quantum information science where the proximity of the ion to control fields is important. As with the common linear five-wire trap, shuttling of ions can be used to spatially separate readout and control regions above a surface trap chip. For instance, when large magnetic field gradients for microwave gates are required \cite{lekitsch2017blueprint, Ospelkaus2011, Mintert2001} the ion can be brought closer to the surface for the duration of a quantum gate while readout beams are kept further away to avoid scattering from the surface and possible charging effects \cite{Wang2011Laser-inducedTraps, Allcock2011a, Harlander2010}. Vertical ion shuttling also has advantages for ion loading, which can be performed further away from the surface to avoid coating the it with Ca atoms. In this work we loaded ions at 150~{\textmu}m height and then moved the ion to the desired height.

Surface traps based on this design may also be used for remote coupling of charged particles via conducting interconnects ~\cite{Daniilidis2009}, where the distance to the coupling element determines the coupling strength. This approach may be interesting for linking ions in separate regions of an extended surface trap for quantum computing.

\section*{Acknowledgments}
This work has been supported by AFOSR through Grant No. FA9550-15-1-0249 and by the NSF Grant No. PHY 1507160. Metal evaporation was performed in the UC Berkeley Marvell Nanofabrication Laboratory. We thank Phillip Schindler and Nikos Daniilidis for insightful discussions. E. U. and M. B.-U. acknowledge support by the NSF Graduate Research Fellowship.

\bibliography{mendeley}

\end{document}